\documentclass{article}

\usepackage{geometry}

\usepackage{epsfig}
\usepackage{amsfonts}

\usepackage{amssymb,amsmath, amsthm}

\def\a{\alpha}
\def\ve{\varepsilon}
\def\G{\Gamma}
\def\no{\nonumber}

\def\le{\left(}
\def\ri{\right)}
\def\L{\lambda}

\begin{document}

\begin{titlepage}
\vskip 1cm
\begin{center}
{\Large \bf  Multi-fold contour integrals of certain ratios \\
\vskip 2mm
of Euler gamma functions from Feynman diagrams: \\
\vskip 2mm
orthogonality of triangles}\\
\vskip 5mm  
Ivan Gonzalez $^{(a)},$  Igor Kondrashuk $^{(b)},$ Eduardo A. Notte-Cuello $^{(c)},$  Ivan Parra-Ferrada $^{(d)}$ \\
\vskip 5mm  
{\it  (a) Instituto de F\'isica y Astronom\'ia, Universidad de Valpara\'iso, \\ Av. Gran Breta\~na 1111,  Valpara\'iso, Chile}  \\
{\it  (b) Grupo de Matem\'atica Aplicada {\rm \&} Grupo de F\'isica de Altas Energ\'ias, \\ 
          Departamento de Ciencias B\'asicas,  Universidad del B\'io-B\'io, Campus Fernando May, \\
          Av. Andres Bello 720, Casilla 447, Chill\'an, Chile} \\
{\it  (c) Departamento de Matem\'aticas, Facultad de Ciencias, Universidad de La Serena, \\
          Av. Cisternas 1200, La Serena, Chile}      \\
{\it  (d) Fakult\"at f\"ur Physik, Universit\"at Bielefeld, Universit\"atsstra\ss e~25, 33615 Bielefeld, Germany} 
\\
\vskip 5mm  
{\it Dedicated to the memory of Sasha Vasil'ev}
\end{center}

\begin{abstract}
We observe a property of orthogonality of the Mellin-Barnes transformation of the triangle one-loop diagrams,
which follows from our previous papers [JHEP {\bf 0808} (2008) 106, JHEP {\bf 1003} (2010) 051, JMP {\bf 51} (2010) 052304]. 
In those papers it has been established that Usyukina-Davydychev functions are invariant with respect to 
Fourier transformation. This has been proved at the level of graphs and also via the Mellin-Barnes transformation. 
We  partially apply to one-loop massless scalar diagram the same trick  in which the  Mellin-Barnes transformation was involved and 
obtain the property of orthogonality of the corresponding MB transforms under integration over contours in two complex planes 
with certain weight. This property is valid in an arbitrary number of dimensions. 
\end{abstract}

\section{Foreword of I.K.}

I met Sasha Vasil'ev for the first time in 2001 in Valparaiso when I came as a postdoctoral fellow to the high energy physics group at the Physics Department of 
UTFSM. At that time Sasha and Irina were professors at the Mathematical Department. He was a wonderful person and a brilliant mathematician. 
He worked in many fields of mathematics. Complex analysis was one of them. My talk is dedicated to an application of complex analysis to the tasks that 
may appear from comparing integrals corresponding to different Feynman diagrams.  My talk is along the lines of paper \cite{Allendes:2012mr}. Sasha was my friend and his pass away is a loss for me.
He was a real human.

\end{titlepage}

\section{Introduction}

Ladder diagrams are important family of Feynman diagrams. The scalar ladder diagrams have been calculated in Refs. \cite{Belokurov:1983km,Usyukina:1992jd,Usyukina:1993ch}  
at the beginning  of nineties in $d=4$ space-time dimensions. 
The result is written in terms of Usyukina-Davydychev (UD) functions \cite{Usyukina:1992jd,Usyukina:1993ch}. 
The calculation was based on the loop reduction technique proposed in Ref. \cite{Belokurov:1983km}.  
With help of this loop reduction technique  any diagram of this family may be reduced to a one-loop massless triangle diagram with a bit modified indices of propagators. 
The trick of loop reduction has been generalized to an arbitrary space-time dimension in Refs. \cite{Gonzalez:2012gu,Gonzalez:2012wk},  however with the index of the rungs different from 1.  

Mellin-Barnes (MB) transforms of the momentum integrals corresponding to ladder diagrams at any loop order  have been studied in Refs. \cite{Allendes:2009bd,Allendes:2012mr,Kniehl:2013dma,Gonzalez:2016pgx}  
in four-dimensional case. In Ref. \cite{Gonzalez:2016pgx} it has been shown that the loop reduction in terms of Mellin-Barnes transforms means the application of the first and  the second Barnes lemmas 
\cite{Barnes-1,Barnes-2}. The main equation of loop reduction technique in terms of MB transforms  may be written explicitly in two lines \footnote{We omit the factor $1/2\pi i$ in front of each contour integral in 
complex plane}
\begin{eqnarray} \label{R1}
\oint_C~dz_2dz_3~D^{(u,v)}[1+\ve_1-z_3,1+\ve_2-z_2,1+\ve_3] 
D^{(z_2,z_3)}[1+\ve_2,1+\ve_1,1+\ve_3] =  \no\\
J\left[ \frac{D^{(u,v-\ve_2)}[1-\ve_1]}{\ve_2\ve_3}
+  \frac{D^{(u,v)}[1+\ve_3]}{\ve_1\ve_2}  + \frac{ D^{(u-\ve_1,v)}[1-\ve_2]}{\ve_1\ve_3}  \right]  
\end{eqnarray}
and is given for the first time in Ref. \cite{Allendes:2012mr}, where $D^{(u,v)}[\nu_1,\nu_2,\nu_3]$ 
is MB transform of the one-loop massless triangle scalar momentum integral,  $\ve_1,\ve_2,\ve_3$ are three complex numbers which satisfy the condition $\ve_1+\ve_2+\ve_3 = 0,$  and 
$J$ is a factor that depends on $\ve_1,\ve_2,\ve_3$ only.

Eq.(\ref{R1}) has a structure similar to decomposition of tensor product in terms of irreducible components.  In the present paper we find an analog of orthogonality condition.  
The key step to this aim is to repeat the proof that has been applied in Ref. \cite{Allendes:2009bd} in order to show invariance of Usyukina-Davydychev functions
with respect to Fourier transformation.  This proof is simple and may be used to show the invariance with respect to Fourier transformation of any three-point Green function \cite{Kondrashuk:2009us}. 
The orthogonality and the decomposition taken together  suggest  that in quantum field theory there are internal integrable structures  for the Green functions.  Integrable structures are usually known for amplitudes 
\cite{Bern:2006ew} but not for  the multi-point Green functions.      

The invariance of UD functions with respect to Fourier transformation is known already for several years \cite{Kondrashuk:2008ec,Kondrashuk:2008xq,Kondrashuk:2009us}. 
The Fourier invariance in these papers has been found  
at the level of graphs. Then, this invariance has been proved via Mellin-Barnes transformation in Ref. \cite{Allendes:2009bd}. In the next sections we collect all the necessary tools 
to prove the orthogonality of the triangle MB transforms  $D^{(u,v)}[\nu_1,\nu_2,\nu_3]$ and find a weight with which this orthogonality appears inside complex integrals over contours.

\section{MB representation of one-loop triangle diagram}

One-loop massless triangle diagram is depicted in Figure 1. It contains three scalar propagators. The $d$-dimensional momenta $p_1$, $p_2$, $p_3$ enter this diagram. 
They are related by  momentum conservation
\begin{eqnarray} \label{conser}
p_1 + p_2 + p_3 = 0.  
\end{eqnarray}
This momentum integral 
\begin{eqnarray*}
J(\nu_1,\nu_2,\nu_3) = \int~Dk~\frac{1}{\left[(k + q_1)^2\right]^{\nu_1} \left[(k + q_2)^2 \right]^{\nu_2}
\left[(k + q_3)^2\right]^{\nu_3}}
\end{eqnarray*}
corresponds to the diagram in Figure 1. The running momentum $k$ is the integration variable. 
Notation is chosen in such a way that the index of propagator $\nu_1$ stands on the line opposite to the vertex of triangle into which the momentum $p_1$ enters. 
\begin{figure}[ht!] 
\centering\includegraphics[scale=0.6]{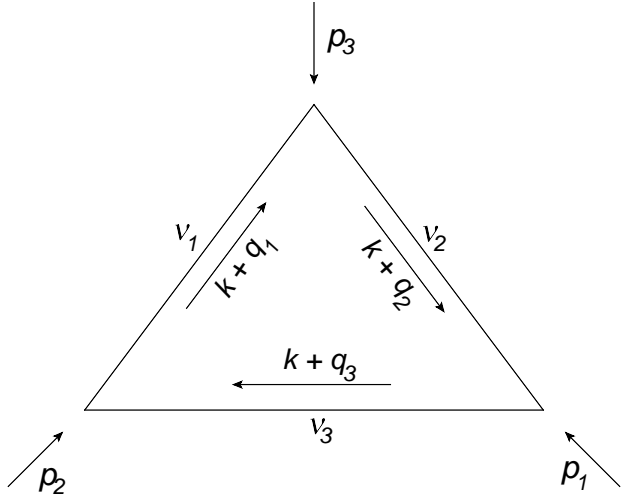}
\caption{\footnotesize  One-loop massless scalar triangle in momentum space}
\label{figure-1}
\end{figure}
The notation $q_1,$ $q_2$ and $q_3$ are taken from Ref.\cite{Usyukina:1992jd}.
It follows from the diagram in Figure \ref{figure-1} and the momentum conservation law that  
\begin{eqnarray*}
p_1 = q_3 - q_2, ~~~ 
p_2 = q_1 - q_3, ~~~
p_3 = q_2 - q_1. 
\end{eqnarray*}
To define the integral measure in momentum space, we use the notation from Ref. \cite{Cvetic:2006iu}
\begin{eqnarray}
Dk \equiv \pi^{-\frac{d}{2}}d^d k.    \label{k-measure}
\end{eqnarray} 
Such a definition of the integration measure in the momentum space helps to avoid powers of $\pi$ in formulas for momentum integrals which will appear in the next sections.  
Mellin-Barnes representation of integral $J(\nu_1,\nu_2,\nu_3)$ may be obtained via Feynman parameters \cite{Usyukina:1992jd} and has the form 
\begin{eqnarray}
J(\nu_1,\nu_2,\nu_3) = \frac{1}{\Pi_{i} \G(\nu_i) \G(d-\Sigma_i \nu_i)} \frac{1}{{(p^2_3)}^{ \Sigma \nu_i -d/2}}
\oint_C dz_2~dz_3 \le\frac{p^2_1}{p^2_3}\ri^{z_2}  \le\frac{p^2_2}{p^2_3}\ri^{z_3} 
\left\{ \G \le -z_2 \ri \G \le -z_3 \ri \right.\no\\
\left. \G \le -z_2 -\nu_2-\nu_3 + d/2 \ri \G \le -z_3-\nu_1-\nu_3 + d/2 \ri 
\G \le z_2 + z_3  + \nu_3 \ri  \G \le  \Sigma \nu_i - d/2 + z_3 + z_2 \ri \right\} \equiv \label{J-arb}\\
\equiv \frac{1}{{(p^2_3)}^{ \Sigma \nu_i -d/2}}\oint_C dz_2~dz_3 x^{z_2}  y^{z_3} 
D^{(z_2,z_3)} [\nu_1,\nu_2,\nu_3]. \no
\end{eqnarray}
We have used here the definition of the Mellin-Barnes transform $D^{(z_2,z_3)}[\nu_1,\nu_2,\nu_3]$ from our Ref. \cite{Allendes:2012mr}, 
\begin{eqnarray}
D^{(z_2,z_3)}[\nu_1,\nu_2,\nu_3] = \frac{ \G \le -z_2 \ri \G \le -z_3 \ri \G \le -z_2 -\nu_2-\nu_3 + d/2 \ri 
\G \le -z_3-\nu_1-\nu_3 + d/2 \ri }
{\Pi_{i} \G(\nu_i) } \no\\
\times 
\frac{ \G \le z_2 + z_3  + \nu_3 \ri  \G \le  \Sigma \nu_i - d/2 + z_3 + z_2 \ri }
{\G(d-\Sigma_i \nu_i)}. \label{J-arb-2}
\end{eqnarray}
The function $D^{(z_2,z_3)}[\nu_1,\nu_2,\nu_3]$ has already appeared in Introduction  in Eq. (\ref{R1}).

\section{Fourier transformation}

The formulas for the Fourier transforms of the massless propagators may be found, for example,  in Ref.~\cite{Allendes:2012mr}. 
The formula we need is 
\begin{eqnarray*}
\int~d^d p ~e^{ip x} \frac{1}{(p^2)^\a} = \pi^{d/2}\frac{\G(d/2-\a)}{\G(\a)}\left(\frac{4}{x^2}\right)^{d/2-\a}, 
\end{eqnarray*}
or, equivalently, 
\begin{eqnarray*}
\frac{1}{(x^2)^{\a}}  = \pi^{-d/2} 4^{-\a}  \frac{\G(d/2-\a)}{\G(\a)}\int~d^d p ~e^{ip x} \frac{1}{(p^2)^{d/2-\a}}. 
\end{eqnarray*}

\section{Star-triangle relation}

For any three vectors $x_1,$ $x_2$, $x_3$  in $d$-dimensional Euclidean space there is so-called ``star-triangle'' relation,   
\begin{eqnarray} \label{eq:uni}
\int Dx \frac{1}{\left[(x_1-x)^2\right]^{\a_1} \left[(x_2-x)^2\right]^{\a_2} \left[(x_3-x)^2\right]^{\a_3}   } = \frac{\G(d/2-\a_1)}{\G(\a_1)}\frac{\G(d/2-\a_2)}{\G(\a_2)}  \times  \no\\
\times\frac{\G(d/2-\a_3)}{\G(\a_3)}\frac{1}{\left[(x_1-x_2)^2\right]^{d/2-\a_3}\left[(x_2-x_3)^2\right]^{d/2-\a_1} \left[(x_1-x_3)^2\right]^{d/2-\a_2}}.
\end{eqnarray}
under the condition of ``uniqueness''  $\a_1 + \a_2 + \a_3 = d$ for the indices in denominator, $\{\a_1,\a_2,\a_3 \} \notin \{d/2+n\}, n \in \mathbb{Z}_{+} $ This relation may be used in the momentum space as well as in the position space.
In the position space we use the same $d$-dimensional measure we used in Eq.(\ref{k-measure}) for the integration in the momentum space,  
\begin{eqnarray*}
Dx \equiv \pi^{-\frac{d}{2}}d^d x 
\end{eqnarray*}
in order to avoid a power of $\pi$ on the r.h.s. of Eq. (\ref{eq:uni}). This appeared to be very useful measure redefinition in the position space too in Refs. \cite{Allendes:2012mr,Gonzalez:2012wk}
in order to develop the loop reduction technique in the position space.  

This star-triangle relation (\ref{eq:uni}) was widely applied to the integrals corresponding to Feynman diagrams in massless field theories. It has been published in Refs.\cite{Unique,Vasiliev:1981dg,Vasil}  
(for the brief review, see Ref.\cite{Kazakov:1984bw}).
The relation has been further developed  to ``stars'' and ``triangles'' with one-step deviation  from unique ``stars'' and ``triangles'' in Refs.\cite{Usyukina:1983gj,Usyukina:1991cp}.  
For example, in Ref.\cite{Usyukina:1983gj} a loop reduction effect has been discovered due to uniqueness method  in a special limit of the indices of diagrams.  
Later, in Refs.\cite{Belokurov:1983km,Usyukina:1991cp} the loop reduction technique has been established without this special limit due to uniqueness method, too. 
The key point of the method has been published in Ref.\cite{Usyukina:1991cp}. 
All the details of the loop reduction technique are given in Ref. \cite{Allendes:2012mr} for $d=4$ case and in Ref. \cite{Gonzalez:2012gu,Gonzalez:2012wk} for an arbitrary space-time dimension.

The star-triangle relation (\ref{eq:uni}) is a basement for the uniqueness method.  It may be proved in many ways. We need in the present paper a proof via the Mellin-Barnes transformation 
\footnote{We do not know explicit reference to this proof of star-triangle relation via the MB transformation. I.K. explained this proof via MB transformation in his lectures on QFT at UdeC, Chile in May of 2009}.  
We assume in the rest of the paper the concise notation of Ref.~\cite{Cvetic:2006iu},  where $[Ny]= (x_N - y)^2$ and analogously for $[Nz],$ and $[yz] = (y-z)^2$, 
that is, $N=1,2,3$ stands for $x_N=x_1,x_2,x_3$, respectively, which are the external points of the triangle diagram in Figure \ref{figure-1} in the position space. 
According to Eq.(\ref{J-arb}) we may write in our notation 
\begin{eqnarray*} 
\int Dx \frac{1}{\left[(x_1-x)^2\right]^{\a_1} \left[(x_2-x)^2\right]^{\a_2} \left[(x_3-x)^2\right]^{\a_3}   } \equiv \int Dx \frac{1}{[x1]^{\a_1} [x2]^{\a_2} [x3]^{\a_3} } \\
= \frac{1}{\Pi_{i} \G(\a_i) \G(d-\Sigma_i \a_i)} \frac{1}{[12]^{ \Sigma \a_i -d/2}}
\oint_C dz_2~dz_3 \le\frac{[23]}{[12]}\ri^{z_2}  \le\frac{[31]}{[12]}\ri^{z_3} 
\left\{ \G \le -z_2 \ri \G \le -z_3 \ri \right.\no\\
\left. \G \le -z_2 -\a_2-\a_3 + d/2 \ri \G \le -z_3-\a_1-\a_3 + d/2 \ri 
\G \le z_2 + z_3  + \a_3 \ri  \G \le  \Sigma \a_i - d/2 + z_3 + z_2 \ri \right\} = \\
= \frac{1}{\Pi_{i} \G(\a_i) \G(0)} \frac{1}{[12]^{d/2}}
\oint_C dz_2~dz_3 \le\frac{[23]}{[12]}\ri^{z_2}  \le\frac{[31]}{[12]}\ri^{z_3} 
\left\{ \G \le -z_2 \ri \G \le -z_3 \ri  \G \le -z_2 + \a_1 - d/2 \ri  \right.\no\\
\left.  \G \le -z_3+\a_2 - d/2 \ri \G \le z_2 + z_3  + \a_3 \ri  \G \le d/2 + z_3 + z_2 \ri \right\}.
\end{eqnarray*}
In the denominator we have $\G(0)$ and this means we need to have the same factor in the numerator. Otherwise, the result would be zero. This may happen only for the residues 
$z_2 = \a_1-d/2$  and  $z_3 = \a_2-d/2,$  because only for these residues  the factor $\G \le z_2 + z_3  + \a_3\ri$ is equal to   $\G \le 0\ri$ and cancels  $\G \le 0\ri$ in the denominator.
Any other residues produced by Gamma functions with negative sign of arguments will have a vanishing contribution due to $\G \le 0\ri$ in the denominator. Thus, 
\begin{eqnarray*} 
\frac{1}{\Pi_{i} \G(\a_i) \G(0)} \frac{1}{[12]^{d/2}}
\oint_C dz_2~dz_3 \le\frac{[23]}{[12]}\ri^{z_2}  \le\frac{[31]}{[12]}\ri^{z_3} 
\left\{ \G \le -z_2 \ri \G \le -z_3 \ri  \G \le -z_2 + \a_1 - d/2 \ri  \right.\no\\
\left.  \G \le -z_3+\a_2 - d/2 \ri \G \le z_2 + z_3  + \a_3 \ri  \G \le d/2 + z_3 + z_2 \ri \right\} = 
\end{eqnarray*}
\begin{eqnarray*} 
\frac{1}{\Pi_{i} \G(\a_i) } \frac{1}{[12]^{d/2}} \le\frac{[23]}{[12]}\ri^{\a_1-d/2}  \le\frac{[31]}{[12]}\ri^{\a_2-d/2}
\G \le d/2 - \a_1 \ri \G \le d/2 -\a_2 \ri   \G \le d/2 - \a_3 \ri  = \\
\frac{1}{\Pi_{i} \G(\a_i) } \frac{\G \le d/2 - \a_1 \ri \G \le d/2 -\a_2 \ri   \G \le d/2 - \a_3 \ri}{[12]^{d/2-\a_3}[23]^{d/2-\a_1}[31]^{d/2-\a_2}}.
\end{eqnarray*}
This proves the star-triangle relation of Eq. (\ref{eq:uni}).

\section{Fourier invariance}

As we have mentioned in Introduction, the proof of Fourier invariance of UD functions has been given in Refs. \cite{Kondrashuk:2008ec,Kondrashuk:2008xq,Kondrashuk:2009us}
at the level of graphs. Later, it has been proved via Mellin-Barnes transformation in Ref. \cite{Allendes:2009bd}. The method of this proof of Ref. \cite{Allendes:2009bd} will be 
the key tool in the next section, in which the main result of the paper is obtained.  We will briefly repeat the proof  of Ref. \cite{Allendes:2009bd} in this section. 

The Usyukina-Davydychev functions  $\Phi^{(n)}(x,y)$ are
functions of two variables. They are the result of calculation of the integrals corresponding to triangle ladder diagrams \cite{Usyukina:1992jd,Usyukina:1993ch}.  
The explicit form of the UD functions $\Phi^{(n)}$ is given in the same Refs. \cite{Usyukina:1992jd,Usyukina:1993ch}
\begin{eqnarray}
\Phi^{(n)}\le x,y\ri = -\frac{1}{n!\L}\sum_{j=n}^{2n}\frac{(-1)^j j!\ln^{2n-j}{(y/x)}}{(j-n)!(2n-j)!}\left[{\rm Li_j}
\le-\frac{1}{\rho x} \ri - {\rm Li}_j(-\rho y)\right], \label{explicit}
\end{eqnarray}
\begin{eqnarray*}
\rho = \frac{2}{1-x-y+\L}, ~~~~ \L = \sqrt{(1-x-y)^2-4xy}.
\end{eqnarray*}
In Refs. \cite{Kondrashuk:2008ec,Kondrashuk:2008xq} Fourier invariance property for the UD functions, that is,  
\begin{eqnarray} \label{inv}
\frac{1}{[31]^2} \Phi^{(n)}\le \frac{[12]}{[31]},\frac{[23]}{[31]}\ri = 
\frac{1}{(2\pi)^4}\int~d^4p_1d^4p_2d^4p_3 ~ \delta(p_1 + p_2 + p_3) \times\no\\
\times e^{ip_2x_2} e^{ip_1x_1} e^{ip_3x_3} \frac{1}{(p_2^2)^2} \Phi^{(n)}\le \frac{p_1^2}{p_2^2},\frac{p_3^2}{p_2^2}\ri
\end{eqnarray}
has been found at the level of graphs. As it has been mentioned in Ref.\cite{Kondrashuk:2008ec}, a hint for such a kind of relation (\ref{inv}) has appeared from the explicit calculation 
of a Green function in Ref. \cite{Cvetic:2007ds}. 
To prove  formula  (\ref{inv}) via MB transformation means to substitute in Eq. (\ref{inv}) the UD function as 
\begin{eqnarray*}
\Phi^{(n)}\le x,y\ri = \oint_C dz_2dz_3 x^{z_2} y^{z_3} {\cal M}^{(n)}\le z_2,z_3\ri,
\end{eqnarray*}
where ${\cal M}^{(n)}\le z_2,z_3\ri$  is the MB transformation of the UD functions $\Phi^{(n)}\le x,y\ri.$ The explicit form of 
${\cal M}^{(n)}\le z_2,z_3\ri$ can be found in Refs.~\cite{Allendes:2012mr,Kniehl:2013dma}. We do not need any use of the explicit
form of those MB transforms. The proof of the Fourier invariance (\ref{inv}) is 
\begin{eqnarray*}
\frac{1}{(2\pi)^4}\int~d^4p_1d^4p_2d^4p_3 ~ \delta(p_1 + p_2 + p_3) 
e^{ip_2x_2} e^{ip_1x_1} e^{ip_3x_3} \frac{1}{(p_2^2)^2} \Phi^{(n)}\le \frac{p_1^2}{p_2^2},\frac{p_3^2}{p_2^2}\ri \\
= \frac{1}{(2\pi)^8}\int~d^4p_1d^4p_2d^4p_3d^4x_5 e^{ip_2(x_2-x_5)} e^{ip_1(x_1-x_5)} e^{ip_3(x_3-x_5)} 
\frac{1}{(p_2^2)^2} \Phi^{(n)}\le \frac{p_1^2}{p_2^2},\frac{p_3^2}{p_2^2}\ri \\
= \frac{1}{(2\pi)^8}\int d^4p_1d^4p_2d^4p_3d^4x_5\oint_C dz_2dz_3 ~ 
\frac{e^{ip_2(x_2-x_5)} e^{ip_1(x_1-x_5)} e^{ip_3(x_3-x_5)}}{(p_2^2)^{2+z_2+z_3} (p_1^2)^{-z_2} (p_3^2)^{-z_3}}  
{\cal M}^{(n)}\le z_2,z_3\ri  \\
= \frac{(4\pi)^6}{(2\pi)^8}\int d^4x_5 \oint_C dz_2dz_3 ~ 
\frac{\G(-z_2-z_3)}{\G(2+z_2+z_3)} \frac{\G(2+z_2)}{\G(-z_2)} \frac{\G(2+z_3)}{\G(-z_3)}\\ 
\times\frac{2^{2z_2+2z_3-2(2+z_2+z_3)}{\cal M}^{(n)}\le z_2,z_3\ri}{[25]^{-z_2-z_3}[15]^{2+z_2} [35]^{2+z_3} }  = \\
= \oint_C~dz_2dz_3 ~ \frac{{\cal M}^{(n)}\le z_2,z_3\ri}{[12]^{-z_3}[23]^{-z_2}[31]^{2+z_2+z_3} }  
= \frac{1}{[31]^2} \Phi^{(n)}\le \frac{[12]}{[31]},\frac{[23]}{[31]}\ri
\end{eqnarray*}
As it has been remarked in Ref.\cite{Kondrashuk:2009us} the same property is valid for any other three-point scalar Green function in the massless field theory 
because the explicit form of the MB transforms  ${\cal M}^{(n)}\le z_2,z_3\ri$ does not play any role in this proof.

\section{Orthogonality of MB transforms of triangles}

In this Section we establish an orthogonality condition for the MB transforms  $D^{(u,v)}[\nu_1,\nu_2,\nu_3]$  of the one-loop massless scalar diagram. All the content of the previous sections 
will be used in the calculations of this section, however, the key step is to repeat the way used in the previous section in order to show the invariance of three-point
functions with respect to Fourier transformation. We will find a weight with which this orthogonality appears inside complex integrals over contours in two complex planes. 
\begin{figure}[ht!] 
\centering\includegraphics[scale=0.6]{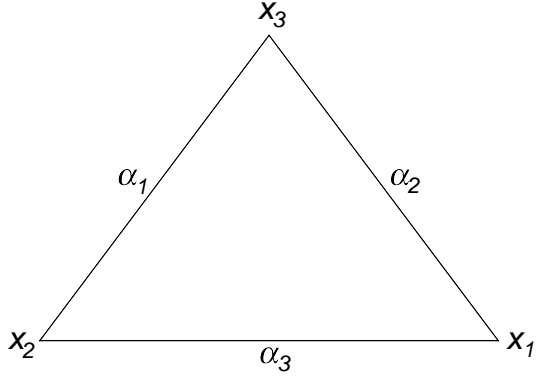}
\caption{\footnotesize  One-loop massless scalar triangle in position space}
\label{figure-2}
\end{figure}
One-loop diagram in the position space is depicted in Figure \ref{figure-2}. It does not contain any integration. Integration in the position space should be done over internal points only (See Ref. \cite{Allendes:2012mr}),
however this diagram does not have internal points at all. The point $x_1,$  $x_2,$  $x_3$ are external points. In the momentum space representation given in Figure \ref{figure-1} the momenta
$p_1,$ $p_2$ and $p_3$ enter.  According to our notation, we write for the diagram in Figure \ref{figure-2}
\begin{eqnarray*}
\frac{1}{[12]^{\a_3}[23]^{\a_1}[31]^{\a_2}} = \frac{  \pi^{-3d/2} 4^{-\Sigma_i \a_i}  \G(d/2-\a_1)\G(d/2-\a_2) \G(d/2-\a_3)}{\G(\a_1)\G(\a_2)\G(\a_3)} \times\\
\int¬dq_1dq_2dq_3 \frac{e^{iq_3(x_1-x_2)} e^{iq_1(x_2-x_3)} e^{iq_2(x_3-x_1)}}{(q_3^2)^{d/2-\a_3}(q_1^2)^{d/2-\a_1} (q_2^2)^{d/2-\a_2}}  = \\
\frac{\pi^{-3d/2} 4^{-\Sigma_i \a_i}\G(d/2-\a_1)\G(d/2-\a_2) \G(d/2-\a_3)}{\G(\a_1)\G(\a_2)\G(\a_3)} 
\int¬dq_1dq_2dq_3 \frac{e^{ix_1(q_3-q_2)} e^{ix_2(q_1-q_3)} e^{ix_3(q_2-q_1)}}{(q_3^2)^{d/2-\a_3}(q_1^2)^{d/2-\a_1} (q_2^2)^{d/2-\a_2}} \\ 
= \frac{\pi^{-3d/2} 4^{-\Sigma_i \a_i}\G(d/2-\a_1)\G(d/2-\a_2) \G(d/2-\a_3)}{\G(\a_1)\G(\a_2)\G(\a_3)} \times\\  
\int¬dp_1dp_2dp_3dq_3\delta(p_1+p_2+p_3)    \frac{e^{ix_1p_1} e^{ix_2p_2} e^{ix_3p_3}}{[q_3^2]^{d/2-\a_3}[(p_2+q_3)^2]^{d/2-\a_1} [(p_1-q_3)^2]^{d/2-\a_2}} \\
= \frac{\pi^{-3d/2} 4^{-\Sigma_i \a_i}\G(d/2-\a_1)\G(d/2-\a_2) \G(d/2-\a_3)}{(2\pi)^d\G(\a_1)\G(\a_2)\G(\a_3)} \times\\  
\int¬dp_1dp_2dp_3dq_3dx_5    \frac{e^{i(x_1-x_5)p_1} e^{i(x_2-x_5)p_2} e^{i(x_3-x_5)p_3}}{[q_3^2]^{d/2-\a_3}[(p_2+q_3)^2]^{d/2-\a_1} [(p_1-q_3)^2]^{d/2-\a_2}} \\
\end{eqnarray*}

\begin{eqnarray*}
= \frac{\pi^{-d} 4^{-\Sigma_i \a_i}\G(d/2-\a_1)\G(d/2-\a_2) \G(d/2-\a_3)}{(2\pi)^d\G(\a_1)\G(\a_2)\G(\a_3)} \times\\  
\int¬dp_1dp_2dp_3dx_5  e^{i(x_1-x_5)p_1} e^{i(x_2-x_5)p_2} e^{i(x_3-x_5)p_3} J(d/2-\a_1,d/2-\a_2,d/2-\a_3) \\
= \frac{\pi^{-d} 4^{-\Sigma_i \a_i}\G(d/2-\a_1)\G(d/2-\a_2) \G(d/2-\a_3)}{(2\pi)^d\G(\a_1)\G(\a_2)\G(\a_3)} \times\\
\int¬dp_1dp_2dp_3dx_5 \oint_C~dz_2dz_3  \frac{e^{i(x_1-x_5)p_1} e^{i(x_2-x_5)p_2} e^{i(x_3-x_5)p_3}} {(p_3^2)^{d-\Sigma_i \a_i +z_2+z_3}(p_1^2)^{-z_2} (p_2^2)^{-z_3} } 
D^{(z_2,z_3)}[d/2-\a_1,d/2-\a_2,d/2-\a_3] \\
= \frac{\pi^{d/2} 4^{d/2}\G(d/2-\a_1)\G(d/2-\a_2) \G(d/2-\a_3)}{(2\pi)^d\G(\a_1)\G(\a_2)\G(\a_3)} \times\\  
\int¬dx_5 \oint_C~dz_2dz_3  \frac{D^{(z_2,z_3)}[d/2-\a_1,d/2-\a_2,d/2-\a_3]} {[35]^{\Sigma_i \a_i -z_2-z_3-d/2}[15]^{d/2+z_2} [25]^{d/2+z_3} } \times\\
\frac{\G(\Sigma_i \a_i -z_2-z_3-d/2 )\G(d/2 + z_2) \G(d/2+ z_3)}{\G(d-\Sigma_i \a_i +z_2 + z_3 )\G(-z_2 )\G(-z_3)} = \\
= \frac{\G(d/2-\a_1)\G(d/2-\a_2) \G(d/2-\a_3)}{\G(\a_1)\G(\a_2)\G(\a_3)} \times\\  
\frac{1}{[12]^{\Sigma_i \a_i}}\oint_C~dz_2dz_3dudv \le\frac{[23]}{[12]}\ri^u  \le\frac{[31]}{[12]}\ri^v  D^{(z_2,z_3)}[d/2-\a_1,d/2-\a_2,d/2-\a_3] \times\\ 
D^{(u,v)} [d/2+z_2, d/2+z_3, \Sigma_i \a_i -z_2-z_3-d/2]  \frac{\G(\Sigma_i \a_i -z_2-z_3-d/2 )\G(d/2 + z_2) \G(d/2+ z_3)}{\G(d-\Sigma_i \a_i +z_2 + z_3 )\G(-z_2 )\G(-z_3)}. 
\end{eqnarray*}
The r.h.s. of this equation may be represented in the form  
\begin{eqnarray*}
\frac{1}{[12]^{\a_3}[23]^{\a_1}[31]^{\a_2}} =   
\frac{1}{[12]^{\Sigma_i \a_i}}\oint_C~dudv \le\frac{[23]}{[12]}\ri^u  \le\frac{[31]}{[12]}\ri^v  f(u,v), 
\end{eqnarray*}
from which we may conclude that  function $f(u,v)$ has residues only at points $u=-\a_1$ and $v=-\a_2$   in the complex planes $u$ and $v.$  The function $f(u,v)$ may be written as double
integral 
\begin{eqnarray*}
f(u,v) =  \frac{\G(d/2-\a_1)\G(d/2-\a_2) \G(d/2-\a_3)}{\G(\a_1)\G(\a_2)\G(\a_3)} \times\\  
\oint_C~dz_2dz_3 D^{(z_2,z_3)}[d/2-\a_1,d/2-\a_2,d/2-\a_3] D^{(u,v)} [d/2+z_2, d/2+z_3, \Sigma_i \a_i -z_2-z_3-d/2]  \\
\frac{\G(\Sigma_i \a_i -z_2-z_3-d/2 )\G(d/2 + z_2) \G(d/2+ z_3)}{\G(d-\Sigma_i \a_i +z_2 + z_3 )\G(-z_2 )\G(-z_3)}. 
\end{eqnarray*}
Thus, we came to a kind of orthogonality condition for the MB transforms, 
\begin{eqnarray} \label{ort}
\oint_C~dudv x^u  y^v  f(u,v) =  x^{-\a_1}y^{-\a_2}.  
\end{eqnarray}

\section{Proof of orthogonality via Barnes lemma}

The orthogonality condition has been found in the previous section in an implicit way. In this section we will prove it explicitly via Barnes lemmas \cite{Barnes-1,Barnes-2} Our results obtained in the 
previous works \cite{Allendes:2012mr,Gonzalez:2016pgx} show that any relation for Mellin-Barnes transforms that follow from Feynman diagrams may be proved via Barnes lemmas. From formula (\ref{J-arb-2})     
we may write 
\begin{eqnarray*}
D^{(z_2,z_3)}[d/2-\a_1,d/2-\a_2,d/2-\a_3] = \G \le  d- \Sigma \a_i  + z_3 + z_2 \ri  \times\\
\times \frac{ \G \le -z_2 \ri \G \le -z_3 \ri \G \le -z_2 +\a_2+\a_3 - d/2 \ri \G \le -z_3+\a_1+\a_3 - d/2 \ri \G \le z_2 + z_3  + d/2-\a_3 \ri}{\G(d/2-\a_1) \G(d/2-\a_2)  \G(d/2-\a_3) \G \le  \Sigma \a_i - d/2 \ri} 
\end{eqnarray*}
and 
\begin{eqnarray*}
D^{(u,v)}[d/2+z_2, d/2+z_3, \Sigma_i \a_i -z_2-z_3-d/2]  =  \G \le  \Sigma \a_i  + u + v \ri  \times\\
\times \frac{ \G \le -u \ri \G \le -v \ri \G \le -u  + z_2 - \Sigma \a_i + d/2   \ri \G \le  -v  + z_3 - \Sigma \a_i + d/2    \ri \G \le u + v  - z_2 - z_3 +  \Sigma \a_i - d/2 \ri}
{\G(d/2+z_2) \G(d/2+z_3) \G(  \Sigma_i \a_i -z_2-z_3-d/2 ) \G \le  d/2-\Sigma \a_i \ri}.
\end{eqnarray*}
We may write explicitly, 
\begin{eqnarray*}
f(u,v) =  \frac{\G(d/2-\a_1)\G(d/2-\a_2) \G(d/2-\a_3)}{\G(\a_1)\G(\a_2)\G(\a_3)} \times\\  
\oint_C~dz_2dz_3 D^{(z_2,z_3)}[d/2-\a_1,d/2-\a_2,d/2-\a_3] D^{(u,v)} [d/2+z_2, d/2+z_3, \Sigma_i \a_i -z_2-z_3-d/2]  \\
\frac{\G(\Sigma_i \a_i -z_2-z_3-d/2 )\G(d/2 + z_2) \G(d/2+ z_3)}{\G(d-\Sigma_i \a_i +z_2 + z_3 )\G(-z_2 )\G(-z_3)} = \\
\frac{\G(d/2-\a_1)\G(d/2-\a_2) \G(d/2-\a_3)}{\G(\a_1)\G(\a_2)\G(\a_3)} \times\\ 
\oint_C~dz_2dz_3  \G \le  d- \Sigma \a_i  + z_3 + z_2 \ri  \times\\
\times \frac{ \G \le -z_2 \ri \G \le -z_3 \ri \G \le -z_2 +\a_2+\a_3 - d/2 \ri \G \le -z_3+\a_1+\a_3 - d/2 \ri \G \le z_2 + z_3  + d/2-\a_3 \ri}{\G(d/2-\a_1) \G(d/2-\a_2)  \G(d/2-\a_3) \G \le  \Sigma \a_i - d/2 \ri} \\
\times \G \le  \Sigma \a_i  + u + v \ri  \times\\
\times \frac{ \G \le -u \ri \G \le -v \ri \G \le -u  + z_2 - \Sigma \a_i + d/2   \ri \G \le  -v  + z_3 - \Sigma \a_i + d/2    \ri \G \le u + v  - z_2 - z_3 +  \Sigma \a_i - d/2 \ri}
{\G(d/2+z_2) \G(d/2+z_3) \G(  \Sigma_i \a_i -z_2-z_3-d/2 ) \G \le  d/2-\Sigma \a_i \ri} \times\\
\frac{\G(\Sigma_i \a_i -z_2-z_3-d/2 )\G(d/2 + z_2) \G(d/2+ z_3)}{\G(d-\Sigma_i \a_i +z_2 + z_3 )\G(-z_2 )\G(-z_3)} = \\
\frac{\G \le  \Sigma \a_i  + u + v \ri  \G \le -u \ri \G \le -v \ri } {\G(\a_1)\G(\a_2)\G(\a_3) \G \le  d/2-\Sigma \a_i \ri\G \le  \Sigma \a_i - d/2 \ri} \times\\ 
\oint_C~dz_2dz_3 \G \le -z_2 +\a_2+\a_3 - d/2 \ri \G \le -z_3+\a_1+\a_3 - d/2 \ri \G \le z_2 + z_3  + d/2-\a_3 \ri  \\
\times \G \le -u  + z_2 - \Sigma \a_i + d/2   \ri \G \le  -v  + z_3 - \Sigma \a_i + d/2    \ri \G \le u + v  - z_2 - z_3 +  \Sigma \a_i - d/2 \ri. 
\end{eqnarray*}
To calculate this integral we need the first Barnes lemma of Ref. \cite{Barnes-1,Smirnov}. 
\begin{eqnarray*} \label{Barnes-1}
\oint_C~dz ~\G \le \lambda_1 + z \ri  \G \le \lambda_2 + z \ri \G \le \lambda_3 - z \ri \G \le \lambda_4 - z \ri   = 
\frac{\G \le \lambda_1 + \lambda_3 \ri \G \le \lambda_1 + \lambda_4 \ri \G \le \lambda_2 + \lambda_3 \ri \G \le \lambda_2 + \lambda_4 \ri}
{\G \le \lambda_1 + \lambda_2 + \lambda_3 + \lambda_4  \ri },  
\end{eqnarray*}
in which  $\lambda_1, \lambda_2, \lambda_3, \lambda_4$ are complex numbers. They are chosen in a such a way  that on the r.h.s. of Eq. (\ref{Barnes-1}) 
there are no singularities. This lemma will  be applied in the next two integrals. 

We first integrate over the variable $z_2,$
\begin{eqnarray*}
\oint_C~dz_2 \G \le -z_2 +\a_2+\a_3 - d/2 \ri  \G \le z_2 + z_3  + d/2-\a_3 \ri  \G \le -u  + z_2 - \Sigma \a_i + d/2   \ri  \times\\ 
\G \le u + v  - z_2 - z_3 +  \Sigma \a_i - d/2 \ri =  \\
\frac{\G \le z_3 + \a_2 \ri \G \le -u - \a_1 \ri \G \le u + v + \a_1 + \a_2 \ri \G \le v - z_3 \ri}
{\G \le v + \a_2  \ri}. 
\end{eqnarray*}
Then, we have 
\begin{eqnarray*}
f(u,v) = \frac{\G \le  \Sigma \a_i  + u + v \ri  \G \le -u \ri \G \le -v \ri \G \le -u - \a_1 \ri \G \le u + v + \a_1 + \a_2 \ri } {\G(\a_1)\G(\a_2)\G(\a_3) \G \le  d/2-\Sigma \a_i \ri\G \le  \Sigma \a_i - d/2 \ri 
\G \le v + \a_2  \ri} \times\\ 
\oint_C~dz_3 \G \le -z_3+\a_1+\a_3 - d/2 \ri   \G \le  -v  + z_3 - \Sigma \a_i + d/2   \ri   \G \le z_3 + \a_2 \ri \G \le v - z_3 \ri = \\
\frac{\G \le  \Sigma \a_i  + u + v \ri  \G \le -u \ri \G \le -v \ri \G \le -u - \a_1 \ri \G \le u + v + \a_1 + \a_2 \ri } {\G(\a_1)\G(\a_2)\G(\a_3) \G \le  d/2-\Sigma \a_i \ri\G \le  \Sigma \a_i - d/2 \ri 
\G \le v + \a_2  \ri} \times\\ 
\frac{\G \le -v - \a_2 \ri \G \le \Sigma \a_i - d/2 \ri \G \le d/2 -  \Sigma \a_i \ri \G \le v + \a_2 \ri}{\G \le 0  \ri} = \\
\frac{\G \le  \Sigma \a_i  + u + v \ri  \G \le -u \ri \G \le -v \ri \G \le -u - \a_1 \ri \G \le -v - \a_2 \ri \G \le u + v + \a_1 + \a_2 \ri } {\G(\a_1)\G(\a_2)\G(\a_3)\G \le 0  \ri }. 
\end{eqnarray*}

This result for $f(u,v)$ allows us to prove the orthogonality condition (\ref{ort}), 
\begin{eqnarray*} 
\oint_C~dudv x^u  y^v  f(u,v) =  \\
\oint_C~dudv  x^u  y^v  \frac{\G \le  \Sigma \a_i  + u + v \ri  \G \le -u \ri \G \le -v \ri \G \le -u - \a_1 \ri 
\G \le -v - \a_2 \ri \G \le u + v + \a_1 + \a_2 \ri } {\G(\a_1)\G(\a_2)\G(\a_3)\G \le 0  \ri }. 
\end{eqnarray*}
The situation is completely the same as in Section 5 where we have proved the star-triangle relation via the Mellin-Barnes transformation. 
In the denominator we have $\G(0)$ and this means we need to have the same factor in the numerator. Otherwise, the result would be zero. This may happen only for the residues 
$u = -\a_1$  and  $v = -\a_2,$  because only for these residues  the factor $\G \le u + v + \a_1 + \a_2 \ri$  is equal to   $\G \le 0\ri$ and cancels  $\G \le 0\ri$ in the denominator.
Any other residues produce by the Gamma functions with negative sign of arguments will have a vanishing contribution due to $\G \le 0\ri$ in the denominator. Thus, 
\begin{eqnarray*} 
\oint_C~dudv  x^u  y^v  \frac{\G \le  \Sigma \a_i  + u + v \ri  \G \le -u \ri \G \le -v \ri \G \le -u - \a_1 \ri 
\G \le -v - \a_2 \ri \G \le u + v + \a_1 + \a_2 \ri } {\G(\a_1)\G(\a_2)\G(\a_3)\G \le 0  \ri } \\
= x^{-\a_1}y^{-\a_2}. 
\end{eqnarray*}

\section{Conclusion}

In this section we have found the explicit form for the integral of two MB transforms over contours in two complex planes, 
\begin{eqnarray} \label{R2}
\oint_C~dz_2dz_3  ~ \Delta^{(\a_1,\a_2,\a_3)}(z_2,z_3)  D^{(u,v)} [d/2+z_2, d/2+z_3, \Sigma_i \a_i -z_2-z_3-d/2]\times \\  
  D^{(z_2,z_3)}[d/2-\a_1,d/2-\a_2,d/2-\a_3]  = \no
\end{eqnarray}
\begin{eqnarray*}
\frac{\G \le  \Sigma \a_i  + u + v \ri  \G \le -u \ri \G \le -v \ri \G \le -u - \a_1 \ri \G \le -v - \a_2 \ri \G \le u + v + \a_1 + \a_2 \ri } {\G(d/2-\a_1)\G(d/2-\a_2) \G(d/2-\a_3) \G \le 0  \ri },
\end{eqnarray*}
where the weight $\Delta^{(\a_1,\a_2,\a_3)}(z_2,z_3)$ is defined as   
\begin{eqnarray*}
\Delta^{(\a_1,\a_2,\a_3)}(z_2,z_3) =  \frac{\G(\Sigma_i \a_i -z_2-z_3-d/2 )\G(d/2 + z_2) \G(d/2+ z_3)}{\G(d-\Sigma_i \a_i +z_2 + z_3 )\G(-z_2 )\G(-z_3)}.
\end{eqnarray*}
The right hand side should be understood in a generalized sense because the presence of $\Gamma(0)$ in the denominator suggests the integration in two complex planes $u$ and $v.$ 
This formula is valid for any space-time dimensions $d.$

On the other side we have Eq.(\ref{R1}) in which the MB transforms of the one-loop massless scalar integral are involved too,   
\begin{eqnarray*} 
\oint_C~dz_2dz_3~D^{(u,v)}[1+\ve_1-z_3,1+\ve_2-z_2,1+\ve_3] 
D^{(z_2,z_3)}[1+\ve_2,1+\ve_1,1+\ve_3] =  \no\\
J\left[ \frac{D^{(u,v-\ve_2)}[1-\ve_1]}{\ve_2\ve_3}
+  \frac{D^{(u,v)}[1+\ve_3]}{\ve_1\ve_2}  + \frac{ D^{(u-\ve_1,v)}[1-\ve_2]}{\ve_1\ve_3}  \right].
\end{eqnarray*}
This equation has been proved in Ref.\cite{Gonzalez:2016pgx} by using the first and the second Barnes lemmas. This equation is valid for $d=4$ space-time dimensions, however, in an arbitrary 
space-time dimension the analog should exist due to the consideration done in Ref. \cite{Gonzalez:2012wk}.

As we have mention in  Introduction,  the integral relation (\ref{R1}) has a structure similar to decomposition of tensor product in terms of irreducible components, and  the integral 
relation  (\ref{R2}) has a structure similar to  orthogonality condition. This observation suggests that behind MB transforms  $D^{(u,v)}[\nu_1,\nu_2,\nu_3]$  an 
integrable structure may exist \cite{Allendes:2012mr,Kniehl:2013dma,Gonzalez:2016pgx}. It is remarkable that both the relations  (\ref{R1}) and (\ref{R2}) are written for the Green functions, that is, 
the integrable structure should exist for the Green functions, too. Usually, integrable structures are studied for amplitudes \cite{Bern:2006ew}.

\subsection*{Acknowledgments}

The work of I.K. was supported in part 
by Fondecyt (Chile) Grants Nos. 1040368, 1050512 and 1121030, by DIUBB (Chile) Grant Nos.  125009,  GI 153209/C  and GI 152606/VC.  
Also, the work of I.K. is supported by Universidad del B\'\i o-B\'\i o and Ministerio de Educacion (Chile)
within Project No.\ MECESUP UBB0704-PD018.
He is grateful to the Physics Faculty of Bielefeld University for accepting
him as a visiting scientist and for the kind hospitality and the excellent
working conditions during his stay in Bielefeld. 
E.A.N.C.  work was partially supported by project DIULS PR15151, Universidad de La Serena.
The work of I.P.F. was supported in part by Fondecyt (Chile) Grant No.\
1121030 and by Beca Conicyt (Chile) via Doctoral fellowship CONICYT-DAAD/BECAS Chile, 2016/91609937.
This paper is based on the talk of I. K. at ICAMI 2017, San Andres, Colombia, November 27 - December 1, 2017, 
and he is grateful to ICAMI organizers for inviting him. The financial support of I.K. participation in ICAMI 
2017 has been provided by DIUBB via Fapei funding.

\end{document}